\documentclass[11pt]{article}

\usepackage{amsmath}
\usepackage{amssymb}
\usepackage{graphicx}
\usepackage{subfigure}

\def\middlespace {\smallskipamount=5.625pt plus1.5pt minus1.5pt
                  \medskipamount=11.25pt plus3pt minus3pt
                  \bigskipamount=22.5pt plus6pt minus6pt
                  \normalbaselineskip=22.5pt plus0pt minus0pt
                  \normallineskip=1pt
                  \normallineskiplimit=0pt
                  \jot=5.625pt
                  {\def\smallskip {\vskip\smallskipamount}}
                  {\def\medskip   {\vskip\medskipamount}}
                  {\def\bigskip   {\vskip\bigskipamount}}
                  {\setbox\strutbox=\hbox{\vrule
                    height15.75pt depth6.75pt width 0pt}}
                  \parskip 11.25pt
                  \normalbaselines}

\begin{document}

\ \vskip 1.0 in

\begin{center}
 { \Large {\bf Noncommutative Gravity, a `No Strings Attached' }}

\smallskip

{\Large {\bf  Quantum-Classical Duality, and the Cosmological Constant Puzzle}}

\vskip 0.2 in

\smallskip

\bigskip

\bigskip

\bigskip

{{\large
{\bf T. P. Singh\footnote{e-mail address: tpsingh@tifr.res.in} 
} 
}}

\medskip

{\it Tata Institute of Fundamental Research,}\\
{\it Homi Bhabha Road, Mumbai 400 005, India.}\\
\medskip
{\it 22nd March, 2008}
\vskip 0.5cm
\end{center}

\vskip 1.0 in

\begin{abstract}

\noindent There ought to exist a reformulation of quantum mechanics which does not refer to an external
classical spacetime manifold. Such a reformulation can be achieved using the language of
noncommutative differential geometry. A consequence which follows is that the `weakly quantum,
strongly gravitational' dynamics of a relativistic particle whose mass is much greater than Planck mass
is dual to the `strongly quantum, weakly gravitational' dynamics of another particle whose mass is 
much less than Planck mass. The masses of the two particles are inversely related to each other,
and the product of their masses is equal to the square of Planck mass. This duality explains
the observed value of the cosmological constant, and also why this value is nonzero but
extremely small in Planck units.

\vskip 1.0 in

\centerline{\it This essay received the Second Prize in the}
\centerline{\it Gravity Research Foundation Essay Competition, 2008}
     
\end{abstract}

\newpage

\middlespace

\noindent 

\noindent In general relativity, the Schwarzschild radius $R_S=2Gm/c^2$ of a particle of mass $m$ can be written in Planck units as $R_{SP}\equiv R_S/L_{Pl}=2m/m_{Pl}$, where $L_{Pl}$ is Planck length and $m_{Pl} \sim 10^{-5}$ gm is the Planck mass. If the same particle were to be treated, not according to general relativity, but according to relativistic quantum mechanics, then one-half of the Compton wavelength $R_C=h/mc$ of the particle can be written in Planck units as $R_{CP}\equiv R_C/2L_{Pl}= m_{Pl}/2m$.  The fact that the product $R_{SP}R_{CP}=1$ is a universal constant cannot be a coincidence; however it cannot be explained in the existing theoretical framework of general relativity (because herein $h=0$) and quantum mechanics (because herein $G=0$). It points to a deeper picture in which general relativity and quantum mechanics are both limiting cases of a quantum gravity theory, 
wherein $R_{SP}$ and $R_{CP}$ can be defined simultaneously. In this essay we motivate the constancy of the product $R_{SP}R_{CP}$ as a consequence of a quantum-classical duality, which we propose. We then
use this duality to explain why the cosmological constant seems to have a tiny non-zero value in the observed Universe.

\noindent We propose and justify the following quantum-classical duality:

{\it
\noindent The weakly quantum, strongly gravitational dynamics of a particle of mass $m_c\gg m_{Pl}$ is dual to the strongly quantum, weakly gravitational dynamics of a particle of mass $m_q=m_{Pl}^2/m_c\ll m_{pl}$.}

It follows that the dimensionless Schwarzschild radius $R_{SP}$ of $m_c$ is four times the dimensionless
Compton-wavelength $R_{CP}$ of $m_q$. 

The origin of this duality lies in the requirement that there be a reformulation of quantum mechanics
which does not refer to an external classical spacetime manifold \cite{singh}. One should be able to formulate quantum mechanics for a quantum system even if there are no external classical bodies in the
Universe. In such a situation, there is no classical gravitational field available, and in accordance with the Einstein hole argument \cite{singh2} the ever-present quantum gravitational fluctuations destroy the point structure of the underlying classical spacetime manifold. Hence the need for a reformulation.
Standard linear quantum mechanics can then be shown to be a limiting case of a nonlinear theory, with
the nonlinearity becoming important in the vicinity of the Planck mass/energy scale, and negligible otherwise. Furthermore, if the point structure of spacetime is lost, the natural new geometric structure which can take its place is noncommutative geometry. 

In this reformulation, the quantum dynamics of a relativistic particle of mass $m\ll m_{Pl}$
is described as a noncommutative special relativity. In the illustrative 2-d case, the noncommutative
spacetime has the line element
\begin{equation}
d\hat{s}^{2}=\hat{\eta}_{\mu\nu}d\hat{x}^{\mu}d\hat{x}^{\nu}\equiv
d\hat{t}^{2}-d\hat{x}^{2}
+d\hat{t}d\hat{x}-d\hat{x}d\hat{t},
\label{lin}
\end{equation}
and the noncommuting coordinates $\hat{t},\hat{x}$ obey the commutation relations 
\begin{equation}
[\hat{t},\hat{x}]=iL_{Pl}^{2}, \qquad [\hat{p}^{t}, \hat{p}^{x}]=iP_{Pl}^{2}.
\label{commu}
\end{equation}
All noncommutative products are to be understood as star products. Dynamics is described by defining a velocity $\hat{u}^{i}=d\hat{x}^{i}/d\hat{s}$, a momentum 
$\hat{p}^{i}=m\hat{u}^{i}$, and by defining momenta as the gradients of a complex action 
$\hat{S}$, in the generalized Casimir relation
\begin{equation}
(\hat{p}^{t})^{2}-(\hat{p}^{x})^{2} + 
\hat{p}^{t}\hat{p}^{x} - \hat{p}^{x}\hat{p}^{t}  = m^{2},
\label{nce}
\end{equation} 
in the spirit of the Hamilton-Jacobi equation \cite{singh}. It is assumed that the line-element and
the commutation relations are invariant under transformations of noncommuting spacetime coordinates.  If an external classical spacetime $(x,t)$ becomes available, the Klein-Gordon equation of standard linear quantum mechanics is recovered from this reformulation via the correspondence
\begin{equation}
(\hat{p}^{t})^{2}-(\hat{p}^{x})^{2} + 
\hat{p}^{t}\hat{p}^{x} - \hat{p}^{x}\hat{p}^{t}  = ({p}^{t})^{2}-({p}^{x})^{2}
 + i\hbar {\partial p^{\mu}\over \partial x^{\mu}}
\label{nceq}
\end{equation}
and by defining the wave-function as $\psi\equiv e^{iS/\hbar}$.  

When the mass of the particle is comparable to Planck mass, the noncommutative line-element
(\ref{lin}) is modified to the curved noncommutative line-element
\begin{equation}
ds^{2}=\hat{h}_{\mu\nu}d\hat{x}^{\mu}d\hat{x}^{\nu}\equiv
\hat{g}_{tt}d\hat{t}^{2}-\hat{g}_{xx}d\hat{x}^{2}
+\hat\theta[d\hat{t}d\hat{x}-d\hat{x}d\hat{t}].
\label{linc}
\end{equation}
Correspondingly, the Casimir relation (\ref{nce}) is generalized to
\begin{equation}
\label{nceq2}\hat g_{tt}(\hat p^t)^2-\hat g_{xx}(\hat p^x)^2+\hat \theta
\left( \hat p^t\hat p^x-\hat p^x\hat p^t\right) =m^2
\end{equation}
and the correspondence rule (\ref{nceq}) to
\begin{equation}
\label{corr}\hat g_{tt}(\hat p^t)^2-\hat g_{xx}(\hat p^x)^2+\hat \theta
\left( \hat p^t\hat p^x-\hat p^x\hat p^t\right)=g_{tt}({p}^t)^2-g_{xx}({p}%
^x)^2+i\hbar \theta {\frac{\partial p^\mu }{\partial x^\mu }}.
\end{equation}
This leads, in the simplest case, where $\theta$ is a function of $m/m_{Pl}$, to the equation of motion
\begin{equation}
\label{hjcn}\left( {\frac{\partial {S}}{\partial t}}\right) ^2-\left( {\frac{%
\partial {S}}{\partial x}}\right) ^2-i\hbar \theta (m/m_{Pl}) \left( {\frac{\partial ^2S%
}{\partial t^2}}-{\frac{\partial ^2S}{\partial x^2}}\right) =m^2
\end{equation}
which is equivalent to a nonlinear Klein-Gordon equation. The noncommutative metric is assumed to
obey a noncommutative generalization of Einstein equations, with the property that $\theta(m/m_{Pl})$
goes to one for $m\ll m_{Pl}$, and to zero for $m\gg m_{Pl}$. Also, as $\theta(m/m_{Pl})\rightarrow 0$
one recovers classical mechanics, and in the limit $\theta\rightarrow 1$ standard linear quantum mechanics is recovered. In the mesoscopic domain, where $\theta$ is away from these limits and the mass $m$ is comparable to Planck mass, both quantum and gravitational features can be defined simultaneously, and new physics arises. The antisymmetric component $\theta$ of the gravitational field plays a crucial role in what follows. This scenario, although described here for simplicity using the 2-d case, continues
to hold in four dimensions.

The Planck mass demarcates the dominantly quantum domain $m < m_{Pl}$ from the dominantly classical
domain $m > m_{Pl}$ and is responsible for the quantum-classical duality. As is evident from (\ref{hjcn}), the effective Planck's constant is $\hbar\theta(m/m_{Pl})$, going to zero for large masses,
and to $\hbar$ for small masses, as expected. Similarly,  the effective Newton's gravitational constant is $G(1-\theta(m/m_{Pl}))$, going to zero for small masses, and to $G$ for large masses. Thus the parameter space $\theta\approx 1$ is strongly quantum and weakly gravitational, whereas
$\theta\approx 0$ is weakly quantum and strongly gravitational. The Compton wavelength $R_{CP}$ for
a particle of mass $m_q$ gets modified to $R_{CE}\equiv R_{CP}\theta(m_q/m_{Pl})$ and the Schwarzschild radius $R_{SP}$ for a mass $m_c$ gets modified to $R_{SE}\equiv R_{SP}(1-\theta(m_c/m_{Pl}))$. We propose
that the dynamics of a mass $m_q\ll m_{Pl}$ is dual to the dynamics of a mass 
$m_c\gg m_{Pl}$ if $R_{SE}(m_c)=4R_{CE}(m_q)$. This holds if $m_c=m_{Pl}^2/m_q$ and 
\begin{equation} 
\theta(m/m_{Pl}) + \theta(m_{Pl}/m) = 1. 
\label{thetacon}
\end{equation}
If (\ref{thetacon}) holds, the solution for the dynamics for a particle of mass $m_c$ can be obtained
by first finding the solutions of (\ref{hjcn}) for mass $m_q$, and then replacing $\theta(m_q/m_{Pl})$ by $1-\theta(m_{Pl}/m_{q})$, and  finally writing $m_c$ instead of $m_q$, wherever $m_q$ appears. 

We can deduce the functional form of $\theta(m/m_{Pl})$ by noting that the contribution of the symmetric part of the metric, $g_{ik}$,
to the curvature, grows as $m$, whereas the contribution of the antisymmetric part $\theta$ must fall with growing $m$. This implies that $1/\theta$ grows linearly with $m$; thus
\begin{equation}
\frac{1}{\theta(m/m_{Pl})}= a(m/m_{Pl})+b,
\end{equation} 
and $\theta(0)=1$ implies $b=1$; and  we set $a=1$ since this simply defines $m_{Pl}$ as the scaling mass. Hence we get $\theta(m/m_{Pl})=1/(1+m/m_{Pl})$, which satisfies (\ref{thetacon}) and thus establishes the duality.
The mapping $m\rightarrow 1/m$ interchanges the two fundamental length scales in the solution : Compton wavelength
and Schwarzschild radius.

Quantum-classical duality has previously been observed in string theory. Our results suggest
one of two possibilities : (i) such a duality is a property of quantum gravity, independent of string theory; or (ii) we have identified a key physical principle underlying string theory, namely, invariance
of physical laws under general coordinate transformations of noncommuting coordinates. The duality we
observe is holographic, by virtue of the above-mentioned relation $R_{SE}(m_c)=4R_{CE}(m_q)$. Thus, the number of degrees of freedom $N$ that a quantum field associated with the particle $m_q$ possesses (bulk property) should be of the order of the area of the horizon of the dual black hole in Planck units (boundary property), i.e. $N\sim m_{Pl}^2/m_{q}^2$. 

The quantum-classical duality helps understand why there should be a cosmological constant of the order of the observed matter density; the most likely explanation for the observed cosmic acceleration. 
If there is a non-zero cosmological constant term $\Lambda$ in the Einstein equations, of the standard form $\Lambda g_{ik}$, it follows from symmetry arguments that in the noncommutative generalization of gravity, a corresponding term of the form $\Lambda \theta$ should also be present. This latter term vanishes in the macroscopic limit $m\gg m_{Pl}$ but is present in the microscopic limit $m\ll m_{Pl}$. 
However, when $m\ll m_{Pl}$, the effective gravitational constant goes to zero, so $\Lambda$ cannot be sourced by ordinary matter. Its only possible source is the zero-point energy associated with the
quantum particle $m\ll m_{Pl}$. Since this zero-point energy is necessarily non-zero, it follows that 
$\Lambda$ is necessarily non-zero. This same $\Lambda$ manifests itself on cosmological scales,
where $\Lambda g_{ik}$ is non-vanishing, because $g_{ik}$ is non-vanishing, even though $\Lambda\theta$ goes to zero on cosmological scales, because $\theta$ goes to zero. 

The value of $\Lambda$ can be estimated by appealing to the deduced quantum-classical duality.
The total mass in the observable Universe is $m_c\sim c^{3} (GH_0)^{-1}$, where $H_0$ is the present value of the Hubble constant. The mass dual to this $m_c$ is $m_q=m_{Pl}^2/m_c \sim hH_0/c^2$,
and $m_q c^2$ is roughly the magnitude of the zero-point energy.
The associated Compton wavelength $\lambda$ is of the order $h/cm_q \sim cH_0^{-1}$. If we consider the quantum field associated with the mass $m_q$, then by the holographic arguments alluded to above, it has an associated number of degrees of freedom $N$ of the order $(cH_{0}^{-1})^{2}/L_P^2$. The vacuum energy density associated with this quantum field, and hence the value of the cosmological constant, is 
$(m_q c^2) N / \lambda^{3} \sim (cH_0)^2/G$ which is of the order of the observed value of $\Lambda$.
Clearly, nothing in this argument singles out today's epoch; hence it follows that there
is an ever-present $\Lambda$, of the order $(cH)^2/G$, at any epoch, with $H$ being the Hubble constant at that epoch. This solves the cosmic coincidence and fine-tuning problems; and difficulties related to an ever-present $\Lambda$ can be addressed, as has been done by Sorkin \cite{Sorkin}. (See also \cite{paddy}).
 
The holographic value for the allowed number of degrees of freedom plays a crucial role in the argument.
The minimum value of the zero point energy, $m_q c^2 \sim hH_0$, corresponds to a frequency $H_0$, which being the
inverse of the age of the Universe, is a natural minimum frequency (infra-red cut-off). 
This corresponds to a contribution $\Lambda_f$ to the
cosmological constant, per degree of freedom, given by
\begin{equation}
 \Lambda_f = \left(\frac{L_{P}}{cH_{0}^{-1}}\right)^{4} L_{P}^{-2}
\end{equation}
and a corresponding energy density
\begin{equation}
\rho_{f} = \left(\frac{L_{P}}{cH_{0}^{-1}}\right)^{4} \rho_{Pl}.
\end{equation}
We recall that the observed $\Lambda$ and its associated energy density can be written as
\begin{equation}
\Lambda_{obs}= H_{0}^{2}/c^{2} = \left(\frac{L_{P}}{cH_{0}^{-1}}\right)^{2} L_{P}^{-2}
\end{equation}
\begin{equation}
\rho_{\Lambda obs} = c^{2}H_{0}^2/G = \left(\frac{L_{P}}{cH_{0}^{-1}}\right)^{2} \rho_{Pl}
\end{equation}
where $\rho_{Pl}$ is Planck energy density.

One could artificially introduce a cut-off to the total zero point energy of the dual quantum field, for example simply by saying that the maximum allowed frequency is Planck frequency. In this case, the number of degrees of freedom $N$ is $E_{Pl}/E_q$, which is equal to $cH_{0}^{-1}/L_{Pl}$. This gives 
$\Lambda=N\Lambda_f=\left({L_{P}}/{cH_{0}^{-1}}\right)^{3} L_{P}^{-2}$ which does not match with observations.
Now consider what values of $\Lambda$ result from other choices of $N$, by writing $N=(cH_{0}^{-1}/L_{P})^{n}$.
Our deduction has been the holographic value $n=2$, which reproduces the correct $\Lambda$. The choices $n=4$ and $n=3$, which correspond to the four volume and the three volume, give wrong values
of $\Lambda$ (too high), whereas $n=1$ also gives a wrong value of $\Lambda$ (too low).
Put another way, the natural minimum frequency is $\omega_{min}=H_0$. Our choice of $N$ is such that
$\omega_{max} = \omega_{Pl}(\omega_{Pl}/H_{0})$, ($N=\omega_{max}/\omega_{min}$). Thus the maximum
frequency is scaled up from Planck frequency by the same factor by which the minimum frequency is scaled down with respect to Planck frequency. It is also the frequency corresponding to the rest mass of the observed Universe, which is of the order $H_{0}^{-1}$. Thus the UV cut-off is not at Planck energy, but at the observed rest mass of the Universe. With hindsight, it seems rather natural that the observed value of the cosmological constant is reproduced when the infra-red and ultra-violet cut-offs for the zero point energy are taken at the cosmological values
$H_0$ and $H_{0}^{-1}$, respectively. The quantum-classical duality proposed here provides the reason as to why quantum zero point energy contributes to gravity in the first place.

The central thesis of our program has been that there are reasons for having a new formulation of quantum mechanics which does not refer to an external classical spacetime manifold
\cite{singh}. The new formulation yields rich dividends : it can address the quantum measurement
problem \cite{singh2}; provide a deeper understanding of D0-brane dynamics in string theory \cite{singh3};
and as discussed here, it possesses a quantum-classical duality which predicts a small
non-zero value for the cosmological constant which matches with observations. Above all, the new formulation is experimentally falsifiable, because its predictions differ from that
of standard linear quantum mechanics in the mesoscopic domain. 

\bigskip

\noindent {\bf Acknowledgements} : It is a pleasure to thank Kumar S. Gupta, T. Padmanabhan and Aseem Paranjape for useful discussions.

\end{document}